\documentclass[twoside,epsf]{article}
\usepackage{graphicx}
\usepackage{multicol} 
\newcommand{\lascia}[1]{}

\newcount\Mac  \Mac=0  
\newcommand{\ifMac}[2]{\ifnum\Mac=1 #1 \else #2 \fi}

\font\tenrsfs=rsfs10
\font\sevenrsfs=rsfs7
\font\fiversfs=rsfs5
\newfam\rsfsfam
\textfont\rsfsfam=\tenrsfs
\scriptfont\rsfsfam=\sevenrsfs
\scriptscriptfont\rsfsfam=\fiversfs
\def\mathscr#1{{\fam\rsfsfam\relax#1}}

\def\Red  {}

\def\Black{}
\def\Blue {}

\newcommand{\ET}{E\hspace{-1.4ex}/{}_T}
\newcommand{\Emiss}{E\hspace{-1.4ex}/\,}
\newcommand{\pb}{\,{\rm pb}}
\newcommand{\riga}[1]{\noalign{\hbox{\parbox{\textwidth}{#1}}}\nonumber}

\newcommand{\branon}{Y}

\renewcommand{\theequation}{\thesection.\arabic{equation}}

\newcommand{\mysection}[1]{\section{\Blue #1\Black }\setcounter{equation}{0}}

\def\sqr#1#2{{\vbox{\hrule height.#2pt\hbox{\vrule width.#2pt
height#1pt \kern#1pt \vrule width.#2pt}\hrule height.#2pt}}}

\newcommand{\mathbb}[1]{{\rm #1}}

\newcommand{\fig}[1]{~\ref{fig:#1}}

\newcommand{\bAk}[3]{\langle #1|#2|#3\rangle}

\newcommand{\GeV}{\,{\rm GeV}}
\newcommand{\TeV}{\,{\rm TeV}}

\newcommand{\One}{\hbox{1\kern-.24em I}}
\newcommand{\NP}{Nucl. Phys.}
\newcommand{\PRL}{Phys. Rev. Lett.}
\newcommand{\PL}{Phys. Lett.}
\newcommand{\PR}{Phys. Rev.}

\newcommand{\eq}[1]{~(\ref{eq:#1})}

\def\circa#1{\,\raise.3ex\hbox{$#1$\kern-.75em\lower1ex\hbox{$\sim$}}\,}
\makeatletter
\def\art{\@ifnextchar[{\eart}{\oart}}
\def\eart[#1]#2#3#4#5#6{{\rm #2}, {\em #3 \bf #4} {\rm (#6) #5} ({\em #1})}
\def\hepart[#1]#2{{\rm #2, \em#1}}
\newcommand{\oart}[5]{{\rm #1}, {\em #2 \bf #3} {\rm (#5) #4}}

\newcounter{alphaequation}[equation]
\def\thealphaequation{\theequation\hbox to
0.6em{\hfil\alph{alphaequation}\hfil}}
\def\eqnsystem#1{
\def\@eqnnum{{\rm (\thealphaequation)}}
\def\@@eqncr{\let\@tempa\relax \ifcase\@eqcnt \def\@tempa{& & &} \or
  \def\@tempa{& &}\or \def\@tempa{&}\fi\@tempa
  \if@eqnsw\@eqnnum\refstepcounter{alphaequation}\fi
\global\@eqnswtrue\global\@eqcnt=0\cr}
\refstepcounter{equation} \let\@currentlabel\theequation \def\@tempb{#1}
\ifx\@tempb\empty\else\label{#1}\fi
\refstepcounter{alphaequation}
\let\@currentlabel\thealphaequation
\global\@eqnswtrue\global\@eqcnt=0 \tabskip\@centering\let\\=\@eqncr
$$\halign to \displaywidth\bgroup \@eqnsel\hskip\@centering
$\displaystyle\tabskip\z@{##}$&\global\@eqcnt\@ne
\hskip2\arraycolsep\hfil${##}$\hfil& \global\@eqcnt\tw@\hskip2\arraycolsep
$\displaystyle\tabskip\z@{##}$\hfil
\tabskip\@centering&\llap{##}\tabskip\z@\cr}
\def\endeqnsystem{\@@eqncr\egroup$$\global\@ignoretrue} \makeatother

\newcount\n

\begin{document}
\centerline{hep-ph/0007267 \hfill IFUP--TH/2000--23 \hfill SNS-PH/00--13}
\vspace{5mm}

\def\Tcontr#1#2#3#4{\raise#2\rlap{\hskip#1
\hbox{\vrule width 0.4pt height#3}%
\raise#3\hbox{\vrule width #4 height 0.4pt}%
\hbox{\vrule width 0.4pt height#3}}}

\Black
\vspace{0.5cm}
\centerline{\LARGE\bf\Red Collider signals of brane fluctuations}
\medskip\bigskip\Black
\centerline{\large\bf Paolo Creminelli}\vspace{0.2cm}
\centerline{\em Scuola Normale Superiore and INFN, Piazza dei Cavalieri 7, I-56126 Pisa, Italy}
\vspace{3mm}
\centerline{\large\bf Alessandro Strumia}\vspace{0.2cm}
\centerline{\em Dipartimento di Fisica, Universit\`a di Pisa and INFN, Pisa, Italia}
\vspace{1cm}
\Blue
\centerline{\large\bf Abstract}
\begin{quote}\large\indent
Assuming that we live on a non rigid brane with TeV-scale tension,
the scalar fields that control the coordinates of our brane in the extra dimensions
give rise to missing energy signals at high-energy colliders
with a characteristic angular and energy spectrum,
identical to the one due to graviton emission in 6 extra dimensions.
LEP bounds and LHC capabilities are analyzed.

\end{quote}\Black
\vspace{0.5cm}

\mysection{Introduction}
Branes in large extra dimensions offer a new possible explanation of why gravity is much weaker than gauge interactions~\cite{DDG}.
Large extra dimensions give clear experimental signals if larger than $0.03\,{\rm mm}$
or if other particles more detectable than gravitons (i.e.\ `sterile neutrinos'~\cite{nuR5d}) propagate in the extra dimensions.
SN1987a poses strong bounds on both possibilities~\cite{SN1987+graviton,SN1987+nu}.
The missing energy carried away by Kaluza Klein (KK) excitations of the
gravitons can be detected at LHC if $\tilde{M}_d\equiv (2\pi)^{\delta/(2+\delta)} M_d\circa{<} (4\div 7)\TeV$~\cite{GRW}
(where $M_d$ is the reduced Planck mass of the $d=4+\delta$ dimensional theory)
depending on $\delta=(6\div 3)$.
Effective operators of dimension 8 are generated by the heaviest KK gravitons~\cite{GRW}.
It would be nice to have further computable signals of large extra dimensions.


Here we discuss an experimental way of testing if we live on a non rigid brane.
If this is the case there are new scalar fields (`branons') that control the coordinate position of our brane
in the extra dimensions.
Branon interactions with particles on the brane are dictated by general considerations~\cite{Lbrane,Kugo,Lbrane3}.
It seems reasonable to assume that branons are light and cannot be directly detected,
so that they only give rise to missing energy signals.
Such signals 
(for example $e\bar{e}\to \gamma \Emiss$ at LEP1 and LEP2 and $pp\to {\rm jet}\ET$, $pp\to \gamma \ET$ at LHC)
can be computed in terms of only {\em one parameter\/} $f$,
related to the tension of the brane $\tau$ and to the number $\delta$ of branons by $f = \tau^{1/4}/\delta^{1/8}$.
Unfortunately these branon signals have the same energy and angular spectrum as
the ``super-string signal'' produced by KK gravitons propagating in $\delta=6$ extra dimensions\footnote{Like gravitons in $6$ extra dimensions,
branons produce also an attractive long range force proportional to 
$m_1 m_2/(rf)^8$~\cite{Kugo}.}.
The magnitude of the two signals coincide when $ \tau = (\pi\delta/30)^{1/2} M_{10}^4$.
In both cases detectable effects appear only at energies so high that
the effective Lagrangian used to compute them becomes questionable.

\medskip

In section~2 we present the effective branon Lagrangian.
In section~3 we discuss branon missing energy signals,
and comment about the reliability and the relevance of our computation.
In section~4 we study their detectability at high energy colliders.
Explicit expressions for the relevant cross sections are given in appendix~A.

\mysection{The brane Lagrangian}
The minimal brane action is~\cite{Lbrane,Kugo,Lbrane3}
\begin{equation}\label{eq:Lbrane}
{\cal S}_{\rm brane} = \int d^4 x \det e [-\tau + \mathscr{L}_{\rm SM}],
\end{equation}
where $e_\mu^\alpha$ is the induced vierbein and $\mathscr{L}_{\rm SM}$ is the covariant Standard Model (SM) Lagrangian.
We denote by $y_i(x)$ ($i=1,\ldots \delta$) the $\delta$ coordinates of the point $x$ of our brane in the $\delta$ extra dimensions.
Even neglecting bulk gravity, $e$ is non trivial when the brane is warped:
$e_\mu^\alpha = \left(1-{(\partial^\alpha y_i)(\partial_\mu y_i)}\right)^{1/2}$~\cite{Kugo}.
Expanding at first order in $\tau$ and rescaling $\branon_i = y_i\sqrt{\tau} $ 
$${\cal S}_{\rm brane} = \int d^4x\bigg[-\tau + \mathscr{L}_{\rm SM}+ \frac{(\partial_\mu \branon_i)(\partial_\nu \branon_i)}{2} (\delta_{\mu\nu}+
\frac{T_{\mu\nu}^{\rm SM}}{\tau}) +\cdots \bigg]$$
one obtains the flat-space SM Lagrangian $\mathscr{L}_{\rm SM}$, coupled with $\delta$ canonically
normalized scalar fields $\branon_i$ proportionally to the SM energy-momentum tensor
$$T_{\mu\nu}^{\rm SM} = \sum_{e,u,d,\ldots}
\frac{i}{4}(\bar{\Psi}\gamma_{(\mu} D_{\nu)}\Psi - D_{(\mu}\bar{\Psi} \gamma_{\nu)}\Psi) +\sum_{\gamma, G}
(F^a_{\mu\rho}F^a_{\rho\nu}+\frac{\eta_{\mu\nu}}{4} F_{\rho\sigma}^a F_{\rho\sigma}^a) + \cdots$$
Here $D_\mu = \partial_\mu -i g_3  g^a_\mu(x) T^a - i e \gamma_\mu(x) q$,
$F_{\mu\nu}$ are the usual field strength tensors of the photon $\gamma_\mu$ and gluons $g^a_\mu$.
Finally, $\cdots$ denotes the remaining $W,Z$ and higgs contributions.

We now list various different ways in which a
non minimal brane action can differ from the minimal one that we consider.
\begin{itemize}
\item The collective position of $n>1$ coinciding $D$-branes arising in string theory
is described by a non minimal set of branon-like fields.
\item Effects associated with a non-minimal
brane structure (`fat branes') have been studied in~\cite{fat}.
\item The presence of our brane spontaneously breaks also a part of the $d$-dimensional Lorentz symmetry.
However there is no need of introducing more Goldstone fields than the `branons' corresponding to the breaking of translational invariance.
\item If extra dimensions are flat, when bulk gravity is taken into account branons are eaten by the off-diagonal components of the metric
that get a mass of order $f^2/M_{\rm Pl}$~\cite{Kugo}.
We can neglect this mass and, due to the equivalence theorem, describe low-energy interactions in terms of branons.
If instead some extra-dimensions are not flat so that
a shift in the extra dimensions requires energy,
some branons acquire non negligible mass terms~\cite{Lbrane3}.
The main effect of such mass terms would be that only branons lighter than $\sim \sqrt{s}/2$ affect processes at
a given center of mass energy $\sqrt{s}$.
For example bounds from SN1987a~\cite{Kugo} and LEP1 (see below) do not apply if all branons are heavier than $M_Z/2$.
\end{itemize}
Unless otherwise indicated, we assume that extra dimensions have the simplest shape (a product of $\delta$ circles) so that
there are $\delta$ massless branons.
Since we will study experimental signals in which branons manifest as missing energy
(so that the signal is proportional to the number $\delta$ of branons)
for our purposes branon interactions are described by a Lagrangian with only one free parameter $f$
$$\mathscr{L}_{\rm int} = \frac{(\partial_\mu\branon )(\partial_\nu\branon) }{2f^4} T^{\rm
SM}_{\mu\nu},$$ where $\branon$ is a single canonically normalized real scalar field and $f^4 \equiv
\tau/\delta^{1/2}$.

\mysection{Missing energy signals}
We study $e\bar{e}\to \gamma \Emiss$ at LEP and $pp\to \gamma \ET$, $pp\to {\rm jet}\ET$ at LHC.
Explicit expressions for the differential cross sections for all contributing processes and parton subprocesses
\begin{equation}\label{eq:processi}
e\bar{e}\to \gamma \branon \branon ,\qquad
q\bar{q}\to \gamma\branon \branon ,\qquad
q\bar{q}\to g\branon \branon ,\qquad
qg\to q\branon \branon ,\qquad
\bar{q} g\to \bar{q}\branon \branon ,\qquad
gg\to g\branon \branon 
\end{equation}
are listed in the appendix.
Here we show that {\em branons produce the same missing energy signals
as
gravitons propagating in $\delta=6$ extra dimensions}\footnote{Another missing energy signal~\cite{gravitini} with the same energy
dependence and with a different angular spectrum can arise in supersymmetric models from gravitino production,
if supersymmetry is spontaneously broken at TeV energies.}.
In particular, in both cases the signals grow as the 6th power of the collision energy.
The overall normalization of branon and graviton signals is the same if
\begin{equation}\label{eq:equiv}
\tau^2 = \frac{\pi \delta}{30} {M_{10}^8},\qquad\hbox{i.e.}\qquad
f^8 = \frac{\tilde{M}_{10}^8}{1920\pi^5},
\end{equation}
where $\tau$ is the brane tension\footnote{In
string models
the tension of one 4-dimensional supersymmetric $D$ brane moving in the simplest 
6-dimensional compactified space
(a product of 6 circles with equal radii)
is predicted to be $\tau=\tau_1 =  \sqrt{\pi}M_{10}$~\cite{Lbrane}.
Therefore, using eq.\eq{equiv}, in this toy string model
the emission of gravitons give effects 5 times larger than the emission of branons.
In such models $n$ coinciding branes have tension $\tau_n = n\tau_1$,
a gauge group ${\rm U}(n)$ localized on them, and
$n^2$ branons~\cite{Lbrane}.
This gives the hope of obtaining a realistic gauge group.
Branon effects get suppressed by the larger brane tension and
enhanced by the larger number of branons.
However 
$n^2-1$ of the $n^2$ branons are in the adjoint of the ${\rm U}(n)$ gauge group.
No light scalar gluons or scalar $W_\pm$ bosons exist.
These unwanted states become heavy if some force,
maybe generated after supersymmetry breaking,
keeps the $n$ branes together.
The tension of the $n$ bound $D$-branes becomes
$\tau_n = n \tau_1 - $ (some binding energy).
In conclusion,
the `prediction' for the ratio between branon and graviton rates 
gets corrected by unknown model
dependent order one factors.
}, $M_{10}$ is the
reduced Planck mass in 10 dimension,
$\tilde{M}_{10}\equiv (2\pi)^{3/4} M_{10}$ is the phenomenological parameter used in~\cite{GRW} and
$f^8\equiv \tau^2/\delta$ is our phenomenological parameter.

To see this, consider for example the branon process
$e(p_1)\bar{e}(p_2)\to \gamma(q) \branon(k_1) \branon(k_2) $ and the graviton process
$e(p_1)\bar{e}(p_2)\to \gamma(q) G(k) $ (with $k\equiv k_1+k_2$).
We compute the cross section for emitting a photon of energy $q_0 = x\sqrt{s}/2$ and direction $\theta$
with respect to the $e\bar{e}$ beam axis, as measured in the $e\bar{e}$ center of mass frame.
Kinematics fixes $k^2=s(1-x)$.
The differential cross section for producing a photon of given energy and direction in the two cases can be computed as follows.
\begin{itemize}
\item The amplitude for the branon process
can be written as $\mathscr{M} = {\cal T}_{\mu\nu} k_{1\mu} k_{2\nu}/f^4$ where ${\cal T}_{\mu\nu}
\equiv
\bAk{\gamma}{T^{\rm SM}_{\mu\nu}}{e\bar{e}}$ is traceless because we only consider processes involving SM particles with negligible mass.
Decomposing the three body phase space as $d\phi (e\bar{e}\to \gamma \branon \branon ) =
d\phi(\gamma)d\phi(\branon \branon )$ (see eq.\eq{3=2+1}) one has
$$d\sigma(e\bar{e}\to \gamma \branon\branon) = {d\phi(\gamma) \over 2s}\cdot \int
|\mathscr{M}\,|^2\frac{d\phi(\branon \branon )}{2!} =
\frac{x\,dx\,d\cos\theta}{64\pi^2}\cdot
\frac{s^2(1-x)^2}{1920\pi f^8} {\cal T}_{\mu\nu}{\cal T}_{\mu\nu}^*.$$
The integration over the phase space of the two branons has been performed using eq.\eq{kkkk} and noticing that terms
proportional to $k_\mu {\cal T}_{\mu\nu}$ and to ${\cal T}_{\mu\mu}$ vanish since
${\cal T}_{\mu\nu}$ is a conserved traceless tensor.

\item The amplitude for the graviton process 
can be written as $\mathscr{M} = {\cal T}_{\mu\nu} \epsilon_{\mu\nu}/M_{10}^4$ where $\epsilon_{\mu\nu}$ is the polarization tensor 
of the graviton~\cite{GRW}.
Integrating over the orientation of the transverse momentum of the graviton in the 6 extra dimensions,
$\int d^6 k_T = \pi^3 k^4 d(k^2)/2$, one obtains
$$d\sigma(e\bar{e}\to \gamma G) = \int \frac{d^6 k_T}{\tilde{M}_{10}^2}\cdot \frac{d\phi(\gamma G)}{2s}\cdot
\sum_{G~{\rm spin}}|{\cal T}_{\mu\nu}\epsilon_{\mu\nu}|^2 =
\frac{\pi^3 s^3(1-x)^2 dx}{2\tilde{M}_{10}^8}\cdot
\frac{x\,d\cos \theta}{32\pi s}\cdot
{\cal T}_{\mu\nu}{\cal T}_{\mu\nu}^*.$$
The 
sum over the polarizations of the graviton 
has been performed using eq. (43) of~\cite{GRW}, omitting terms that vanish since
${\cal T}_{\mu\nu}$ is a conserved traceless tensor.

\end{itemize}
A comparison of the graviton and branon cross sections gives eq.\eq{equiv}.
The equivalence\eq{equiv} does not hold for processes involving SM particles with non negligible mass\footnote{
The (uncomputable) virtual effects give a simple example of this fact.
For example, let us compare $s$-channel exchange of gravitons and branons with transferred squared momentum $s$.
Tree-level exchange of virtual gravitons in $6$ extra dimensions gives the scattering amplitude~\cite{GRW}
\begin{eqnsystem}{sys:virtuali}
\mathscr{M}=ic_G      (T_{\mu\nu}^{\rm SM}T_{\mu\nu}^{\rm SM} -\frac{1}{8}{T_{\mu\mu}^{\rm SM}
T_{\nu\nu}^{\rm SM}} )&\hbox{where}& c_G=-\frac{1}{\tilde{M}_{10}^8}\int^\Lambda\frac{d^6
k_T}{s-k_T^2}=\frac{\pi^3}{2\tilde{M}_{10}^8}[\frac{\Lambda^4}{2}-s^2\ln(-s)+\cdots]\\
\riga{where the integral runs over transverse graviton momentum.
One loop exchange of virtual branons gives rise to the scattering amplitude}\\
\mathscr{M}=ic_\branon(T_{\mu\nu}^{\rm SM}T_{\mu\nu}^{\rm SM} +\frac{1}{2}{T_{\mu\mu}^{\rm SM}
T_{\nu\nu}^{\rm SM}} )&\hbox{where}& c_\branon=\frac{1}{f^8}\int_0^1 \!\!\! dx \!\! \int^\Lambda
\!\!\! \frac{d^4k}{i(2\pi)^4}
\frac{k^4/24}{[k^2+sx(1-x)]^2}=\frac{5\Lambda^4-s^2\ln(-s)+\cdots}{240 f^8(4\pi)^2}
\end{eqnsystem}
where the integral runs over the loop momentum of the branons.
In both cases we have cut-off divergent integrals in a naive way and we have shown only terms of
order $\Lambda^4$ and $s^2 \ln(-s)$:
\begin{itemize}
\item The $s^2 \ln(-s)$ terms are fixed by the low energy effective theory.
We see that the equivalence\eq{equiv} holds for such terms, but only if the energy-momentum tensor is traceless.
These terms generate an attractive $1/r^8$ force between non relativistic particles~\cite{Kugo} (in the case of gravitons this force of course is
$10$-dimensional gravity).

\item 
The region of integration with $k^2_T,k^2\gg s$ gives rise to effective dimension-8 operators.
In both cases the coefficients of the operators cannot be computed from the low energy effective Lagrangian,
because given by divergent integrals over
transverse graviton momentum and over loop branon momentum, respectively.
\end{itemize}
Such operators could be of experimental interest.
Searches for virtual graviton effects have been performed by
the L3 collaboration~\cite{L3} at LEP2 and by
the D0 collaboration~\cite{D0} at Fermilab.
Using eq.s~(\ref{sys:virtuali}a,b) and
assuming some value for the two $\Lambda$, the resulting `bound' on $M_{10}$ could be directly converted into a `bound' on $f$,
since $T_{\mu\mu}^{\rm SM}$ of the colliding particles is negligible.
}, like Higgs and $Z$ decays.


\bigskip

Up to which values of $\sqrt{s}/f$ can our computation be trusted?
Higher order terms of the expansion in the brane tension
become relevant when $\sqrt{s}\circa{>}(3\div 6) f$ and
give {\em additional\/}  contributions to the missing energy signal,
with $4,6,8,\ldots$ branons in the final state.
As we will see, the missing energy signals emerge from the SM background in a similar range of $\sqrt{s}/f$ values.
Furthermore, the Lagrangian\eq{Lbrane} itself is only
a non-renormalizable effective Lagrangian, whose validity is expected to break down at some unknown energy.
This unknown energy must be smaller than about $\sqrt{s}\circa{<}4f$,
otherwise the 2 body cross section $e\bar{e}\to \branon\branon$ in eq.\eq{eebb} exceeds the unitarity bound
$\sigma \circa{<}1/s$.

\medskip

A more relevant question is:
are branon signals a serious candidate for new physics?
If $f$ is small (for example $f\sim M_Z$)
branons are detectable and kill KK signals~\cite{recoil}, including the ones due to
gravitons.
An optimal solution to the hierarchy problem suggests a sub-TeV value of
the effective cutoff $\Lambda_{\rm UV}$ for quadratically divergent corrections to the higgs mass, since in the SM
$$\delta m_h^2 \approx \delta m_h^2({\rm top})\approx +(0.3 \Lambda_{\rm UV})^2\approx m_h^2 \qquad\hbox{at}\qquad
\Lambda_{\rm UV}\approx 400\GeV.$$
Even assuming that new physics conserve CP, $B$, $L$, $L_i$, $B_i$,
the agreement of precision measurements at the $Z$ pole with SM predictions requires that
various dimensions 6 non-renormalizable operators (NRO) must be suppressed by a scale
$\Lambda_{\rm NRO}$ larger than $(5\div 10)\TeV$~\cite{NRO}.

One possibility is that all scales of new physics are comparable:
if $f\approx M_{d}\approx \Lambda_{\rm UV} \approx \Lambda_{\rm NRO}\circa{>}(5\div 10)\TeV$
there is no conflict with precision measurements but extra dimensions alone do not provide a complete solution to the hierarchy problem,
since $\delta m_h^2({\rm top}) \approx 100 m_h^2$. 
In this case graviton and branon missing energy signals cannot be detected,
and only the lightest string (or whatever) states that interact with SM particles could provide a signal at LHC
(either as NRO or via direct production).

If instead $f\approx M_{d}\approx \Lambda_{\rm UV} < 1\TeV$ extra dimensions really solve the hierarchy problem,
graviton and branon missing energy signals can be detected,
but we do not understand why precision measurements agree with SM predictions.
To know if this is a motivated candidate for new physics
would require a predictive theory of quantum gravity.
String brane models with realistic gauge groups~\cite{BraneModels} could give some useful hint.


\begin{figure}[t]
\begin{center}
\begin{picture}(16,7.8)
\put(0,0){\includegraphics[width=16cm,height=7.5cm]{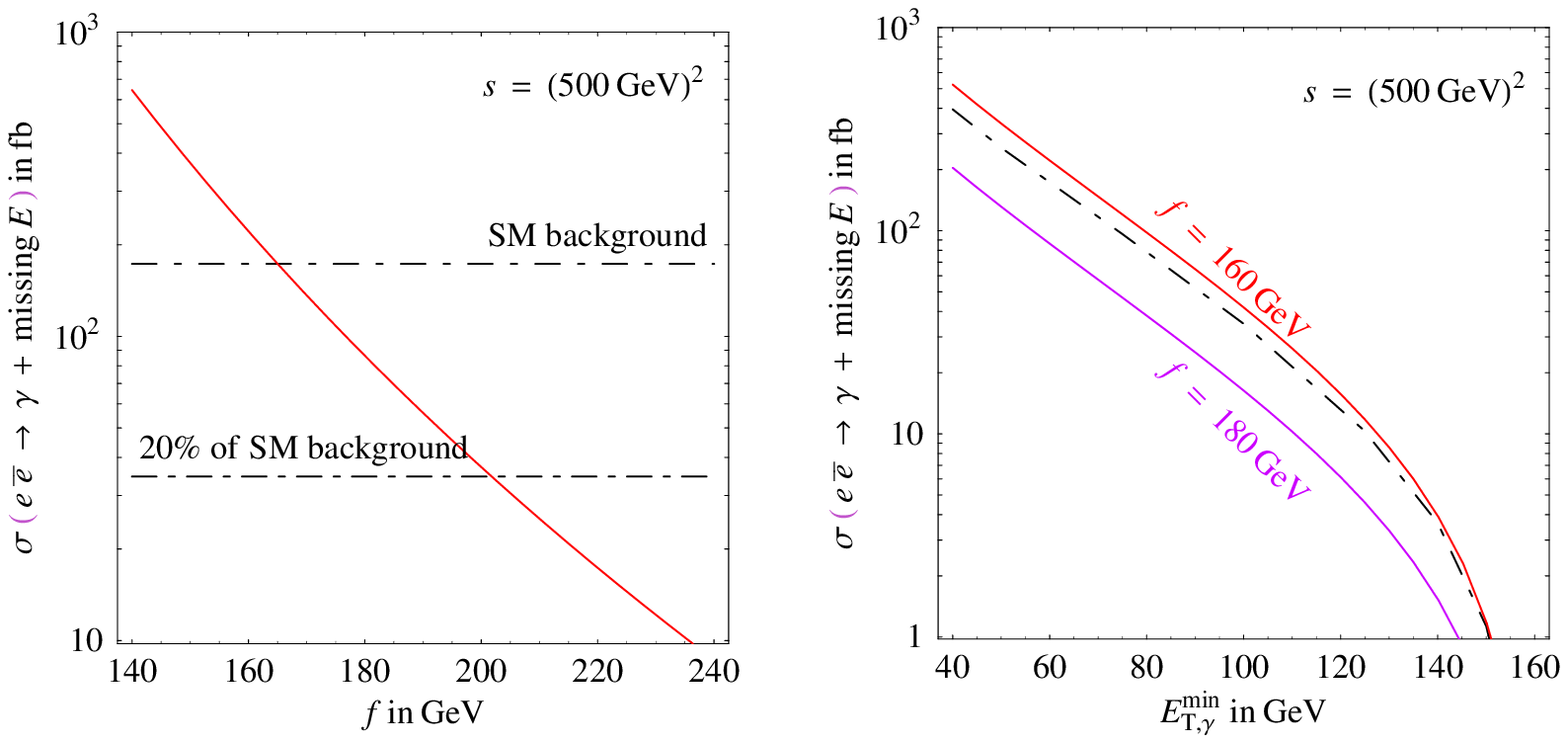}}
\put(3.8,7.6){fig.~\ref{fig:ee500}a}
\put(11.8,7.6){fig.~\ref{fig:ee500}b}
\end{picture}
\caption[SP]{\em Total branon $e\bar{e}\to \gamma\Emiss$ cross section at an $e\bar{e}$ collider
with $\sqrt{s}=500\GeV$.
In fig.~(\ref{fig:ee500}a) we plot the signal with $E_{T\rm \gamma}>E_{T\gamma}^{\rm min}=60\GeV$ and $E_\gamma < 160\GeV$ as function of $f$.
In fig.~(\ref{fig:ee500}b) we plot the signal as function of $E_{T\gamma}^{\rm min}$
for $f=160\GeV$ (upper continuous line) and $f=180\GeV$ (lower continuous line).
The dot-dashed lines represent the SM background.\label{fig:ee500}}
\end{center}\end{figure}

\begin{figure}[t]
\begin{center}
\begin{picture}(16,7.8)
\put(0,0){\includegraphics[width=16cm,height=7.5cm]{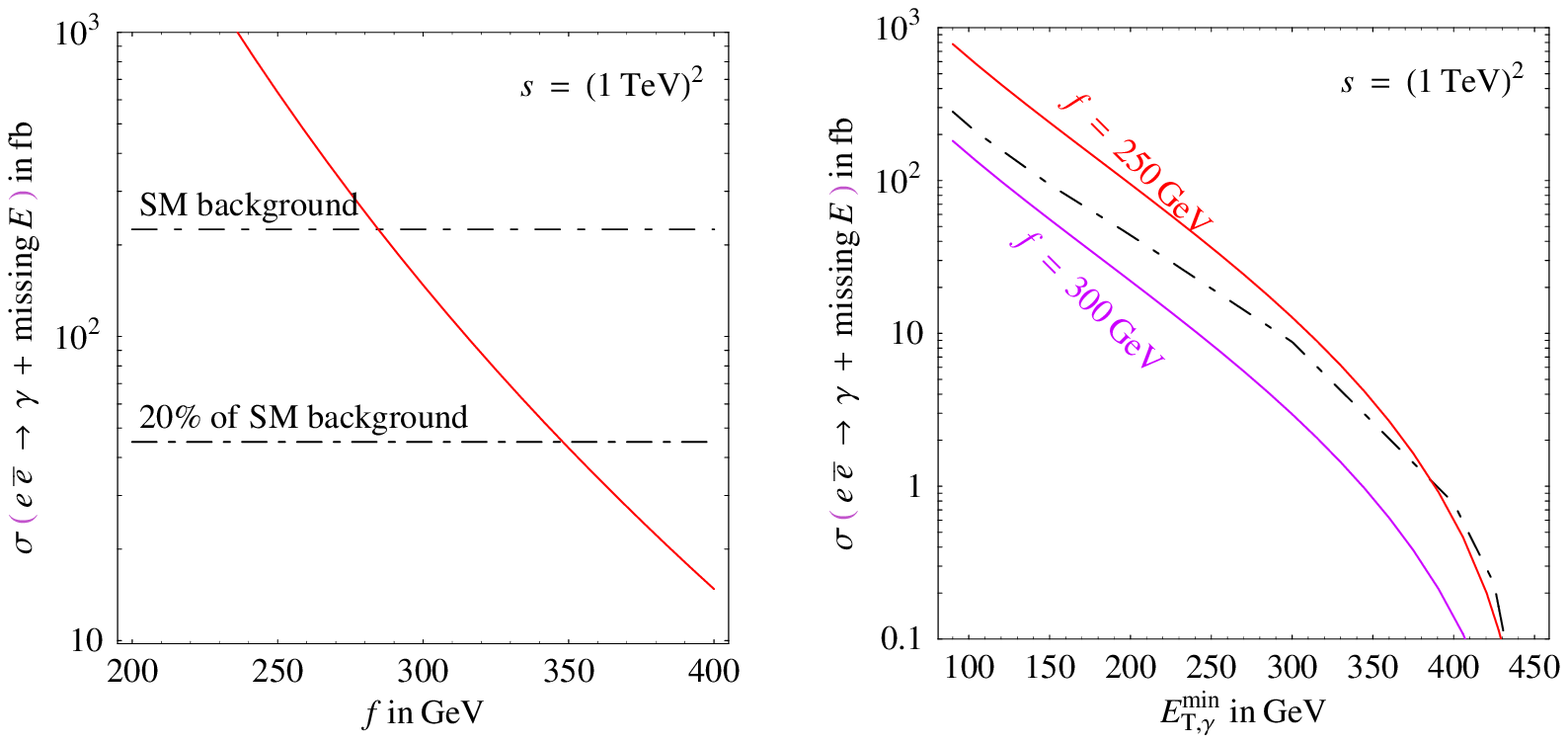}}
\put(3.8,7.6){fig.~\ref{fig:ee1000}a}
\put(11.8,7.6){fig.~\ref{fig:ee1000}b}
\end{picture}
\caption[SP]{\em Total branon $e\bar{e}\to \gamma\Emiss$ cross section at an $e\bar{e}$ collider
with $\sqrt{s}=1\TeV$.
In fig.~(\ref{fig:ee1000}a) we plot the signal with $E_{T\rm \gamma}>E_{T\gamma}^{\rm min}=100\GeV$ and $E_\gamma < 450\GeV$ as function of $f$.
In fig.~(\ref{fig:ee1000}b) we plot the signal as function of $E_{T\gamma}^{\rm min}$
for $f=250\GeV$ (upper continuous line) and $f=300\GeV$ (lower continuous line).
The dot-dashed lines represent the SM background.\label{fig:ee1000}}
\end{center}\end{figure}

\begin{figure}[t]
\begin{center}
\begin{picture}(16,8)
\put(0,0){\includegraphics[width=16cm,height=7.5cm]{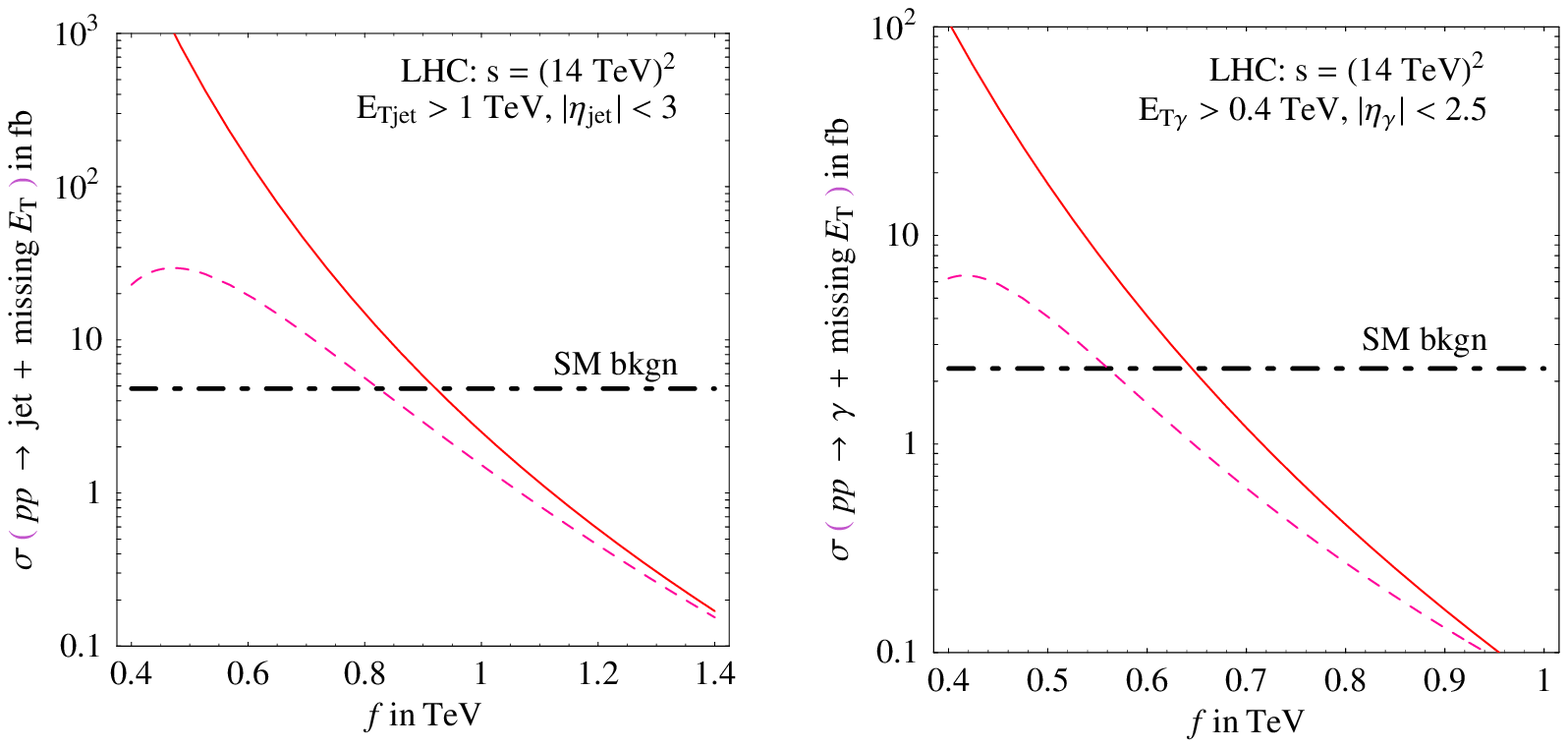}}
\put(3.8,7.6){fig.~\ref{fig:ppf}a}
\put(11.8,7.6){fig.~\ref{fig:ppf}b}
\end{picture}
\caption[SP]{\em The continuous lines show the
total branon cross section for
(\ref{fig:ppf}a) $pp\to {\rm jet}+\ET$ with $E_{T\rm jet}>1\TeV$, $|\eta_{\rm jet}|<3$ and
(\ref{fig:ppf}b) $pp\to \gamma+\ET$ with $E_{T\gamma}>0.4\TeV$, $|\eta_{\gamma}|<2.5$
at LHC with $s=(14\TeV)^2$ as function of $f$.
The dashed line include only the contribution from partonic collisions with 
$\sqrt{\hat{s}} < 2\pi f$.
The horizontal dot-dashed line represents the SM background.
\label{fig:ppf}}
\end{center}\end{figure}

\mysection{Signals at colliders}
Since branons give missing energy signals similar to the gravitino and graviton ones studied in~\cite{gravitini,GRW}
we facilitate a comparison by following these papers for what concerns
tentative experimental cuts and relative backgrounds. 

\subsection{LEP1}
The precision measurements done mostly at LEP1 set a lower bound on $f$
since branons lighter than $M_Z/2$ contribute to the width of the $Z$.
Estimating their contribution as
${\Gamma (Z\to f\bar{f} \branon \branon)}/{\Gamma (Z\to f\bar{f})}\sim 
{(M_Z/f)^8}/{[12(4\pi)^4]}$,
we obtain the bound $f\circa{>} M_Z/2$.
Since branon effects strongly increase with energy, it is not surprising that this bound 
is stronger than the astrophysical bound from the 1987a supernova, $f \circa{>} 10\GeV$~\cite{Kugo}.

\subsection{LEP2}
For the same reason bounds on $e\bar{e}\to \gamma \Emiss$
at LEP2 with $\sqrt{s}=200\GeV$ induce a stronger bound on $f$ than LEP1.
Assuming an integrated luminosity $L = 4\cdot 500\,\pb^{-1}$,
and performing the same cuts on the photon energy and direction as in~\cite{GRW},
the branon signal exceeds the discovery cross section
$\sigma_{\rm discovery} \equiv 5\sqrt{\sigma_{\rm bkgn}/L}= 0.17\pb$ when $f<100\GeV$.
With optimized cuts, LEP collaborations have recently performed a search
for missing energy graviton signals, obtaining bounds on $\tilde{M}_d$.
Using the equivalence\eq{equiv} between branon and graviton signals,
the OPAL bound, $\tilde{M}_{10}>530\GeV$ at $95\%$ confidence level~\cite{OPAL}, can be rescaled to $f> 100\GeV$.

\subsection{Electron colliders}
To explore larger values of $f$ one needs colliders with higher center of mass energy.
In fig.\fig{ee500}a we show, as function of $f$, the total branon $e\bar{e}\to\gamma\Emiss$ cross section
at an $e\bar{e}$ collider with $\sqrt{s}=500\GeV$
integrated over $E_{T\gamma}>60\GeV$ and $E_\gamma<160\GeV$.
$E_\gamma$ is the photon energy, $E_{T\gamma} \equiv E_\gamma \sin\theta $ is the transverse photon energy
and $\Emiss$ is the missing energy, carried away by the branons.
The cut on $E_\gamma$
suppresses the $e\bar{e}\to \gamma Z \to \gamma\nu\bar{\nu}$ SM background.
The dot-dashed line represents the background arising from the SM process $e\bar{e}\to \gamma \nu\bar{\nu}$,
computed with the program {\sc CompHEP}~\cite{CompHEP}.
We see that the signal emerges from the SM background when $0.3\sqrt{s}\circa{>}f$.
At such energies the effective Lagrangian approximation becomes questionable.
A polarization of the electron beams would reduce the background without affecting the signal~\cite{GRW}.
Depending on the acceleration technique, it seems possible to obtain a $80\%$ polarization~\cite{500GeV}.
In fig.\fig{ee500}b we show the missing energy cross section, integrated over $E_\gamma < 160\GeV$ and $E_{T\gamma}>E_{T\gamma}^{\rm min}$
as function of $E_{T\gamma}^{\rm min}$
for $f=160\GeV$ and for $f=180\GeV$.
The dot-dashed line represents the SM background.

Fig.\fig{ee1000} is analogous to fig.\fig{ee500}, but with a higher collider energy: $\sqrt{s}=1\TeV$.
The cuts have been accordingly modified to $E_\gamma<450\GeV$ and $E_{T\gamma}>100\GeV$.

\subsection{LHC}
Next, we study the discovery potential of the approved LHC $pp$ collider with $\sqrt{s}=14\TeV$.
The most promising signal is $pp\to {\rm jet}\ET$ (mostly produced by $qg\to q\branon\branon$).
A less promising signal is $pp\to \gamma \ET$ (produced in $q\bar{q}$ collisions).
The total signal cross sections for these two processes are plotted as function of $f$ and
compared with SM backgrounds in fig.\fig{ppf}a and\fig{ppf}b (continuous lines).
The cuts on the transverse energy $E_T$ and on the pseudorapidity $\eta$ of the jet (fig.\fig{ppf}a) and of the photon (fig.\fig{ppf}b) are specified
in the caption.
The dashed lines show the signal produced including only the contribution from collisions with partonic center of mass energy
$\sqrt{\hat{s}} <2\pi f$.
As discussed in the previous section,
the discrepancy between the continuous and dashed lines indicates
the breakdown of our effective Lagrangian approximation.
There is no good reason for choosing $2\pi f$ rather than another comparable number.
The trustable signal is rapidly reduced below the background if the cut on $\hat{s}$ is lowered.

Almost all the signal is produced by partons that carry $(20\%\div60\%)$ of the proton momentum.
We have used the partonic distribution functions of~\cite{partons}.
With a different choice of partonic distributions, the signal can vary by few $10\%$.

Before LHC, a $p\bar{p}\to{\rm jet}\ET$ signal can be searched at the upgraded Tevatron collider.
However a positive evidence can emerge from the SM background only if $f$ is just above the range excluded by LEP2.

\mysection{Conclusions}
If we really live on a non rigid brane in $\delta$ flat extra dimensions,
the scalar fields that control the coordinates of our brane in the extra dimensions
have low energy interactions with SM particles suppressed by the tension $\tau$ of the brane
and described by a {\em predictive\/} effective non-renormalizable Lagrangian.
We have computed the missing energy signals produced by these
`branon' fields and studied their detectability at high energy colliders.
These signals depend only on one parameter $f\equiv \tau^{1/4}/\delta^{1/8}$.
LEP2 gives the strongest bound on it: $f>100\GeV$.
Branons produce a jet + missing energy signal detectable at LHC if $f\circa{<}900\GeV$.
However, the angular and energy spectrum of the missing energy produced by brane fluctuations in any number of extra dimensions is
identical to the missing energy signal produced by graviton emission in $\delta=6$ extra dimensions.
The two signals have the same magnitude if the brane tension $\tau$ and the 10-dimensional reduced Planck mass
are related by $\tau = \sqrt{\pi \delta/30}M_{10}^4$.
In particular, in both cases the signal rises as the 6th power of the collision center of mass energy:
therefore the energy at which it becomes larger than the SM background is close
to the energy at which the effective Lagrangian approximation becomes questionable. 

\paragraph{Acknowledgments}
We thank Vincenzo Napolano and Riccardo Rattazzi for useful discussions.

\appendix

\mysection{Computation of the cross sections}
When computing cross sections for processes like $e(p_1)\bar{e}(p_2)\to \gamma(q) \branon(k_1)\branon(k_2)$
it is useful to decompose the three body phase space as
\begin{equation}\label{eq:3=2+1}
d\phi^{(3)} (P\to q + k_1 + k_2) = \frac{d^3 q}{2q_0(2\pi)^3} d\phi^{(2)}(k\to k_1+k_2),
\qquad P\equiv p_1+p_2,\qquad
k\equiv k_1+k_2.
\end{equation}
Integrals over the two body phase space of the couple of massless branons can be easily performed
using
\begin{eqnarray}\label{eq:kkkk}
\int d\phi^{(2)}~k_{1\mu} k_{1 \nu} k_{2\rho} k_{2\sigma}&=&  \frac{k_\mu k_\nu k_\rho k_\sigma}{240\pi} +\frac{k^4}{1920\pi} 
(\eta_{\mu\nu} \eta_{\rho\sigma} + \eta_{\mu\rho}\eta_{\nu\sigma}+\eta_{\mu\sigma}\eta_{\nu\rho})+\\
&&- \nonumber
\frac{k^2}{320\pi}(
k_{\mu} k_{\nu} \eta_{\rho\sigma}+
k_{\rho} k_{\sigma} \eta_{\mu\nu})+
\frac{k^2}{480\pi}
(k_{\mu} k_{\rho} \eta_{\nu\sigma}+
k_{\nu} k_{\sigma} \eta_{\mu\rho}+
k_{\mu} k_{\sigma} \eta_{\rho\nu}+
k_{\rho} k_{\nu} \eta_{\mu\sigma}). \nonumber
\end{eqnarray}
We now list the differential cross sections for all processes\eq{processi}.
All cross sections are 
averaged over spins of initial particles and summed over spins of final particles.

\begin{figure}[t]
\begin{center}
\includegraphics[width=14cm]{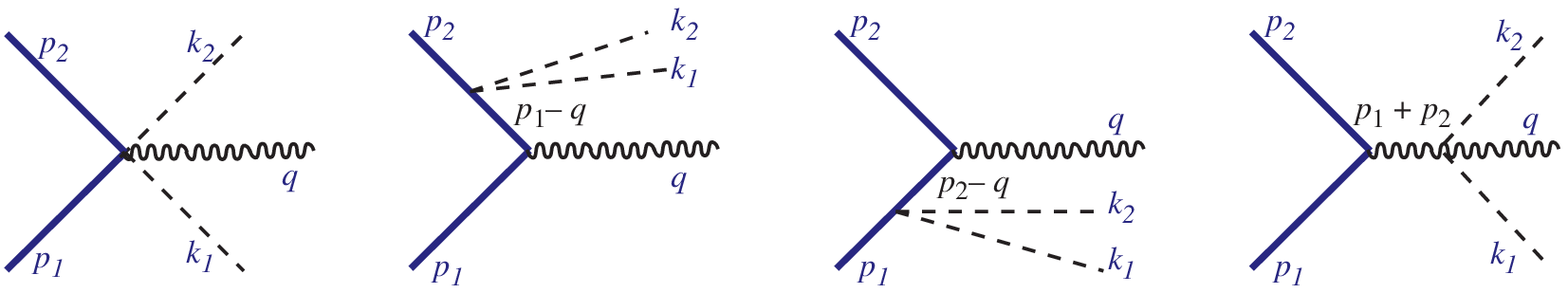}
\caption[SP]{\em The Feynman graphs that contribute to $e\bar{e}\to\gamma \branon\branon$.\label{fig:Feyn}}
\end{center}\end{figure}

\begin{enumerate}
\item The $e\bar{e}\to \gamma \branon\branon$ (or $\mu\bar{\mu}\to \gamma\branon\branon$) cross section is
\begin{equation}\label{eq:eebb}
\frac{d\sigma(e\bar{e}\to\gamma\branon\branon)}{dx\,d\cos\theta} =
\sigma_0 \frac{\alpha_{\rm em}}{4\pi} F_{e\bar{e}},\qquad \hbox{where}\qquad
\sigma_0 \equiv \sigma(e\bar{e}\to\branon\branon)=\frac{s^3 }{30720\pi f^8}.
\end{equation}
The function $F_{e\bar{e}}$ is given at the end of the appendix in terms of the photon energy $E_\gamma = x \sqrt{s}/2$ 
and of the photon direction with respect to beam axis
in the center of mass frame.
The relevant Feynman diagrams are shown in fig.\fig{Feyn}.

\item
The $q\bar{q}\to \gamma \branon\branon$ and $q\bar{q}\to g \branon\branon$ cross sections are
$$\frac{d\sigma(q\bar{q}\to\gamma\branon\branon)}{dt\,d(k^2)} =
\sigma_0 \frac{q_q^2 \alpha_{\rm em}}{12\pi} F_{q\bar{q}},\qquad
\frac{d\sigma(q\bar{q}\to G\branon\branon)}{dt\,d(k^2)} =
\sigma_0 \frac{4}{3}\frac{\alpha_{3}}{12\pi} F_{q\bar{q}}
$$
where $q_q$ is the electric charge of the colliding quark.
The function $F_{q\bar{q}}$ (related in a simple way to $F_{e\bar{e}}$) 
and all functions relative to parton processes are given at the end of the appendix in terms of the invariants $s\equiv (p_1+p_2)^2$,
$t\equiv (p_1-q)^2$, $k^2\equiv (k_1+k_2)^2$
and $u\equiv (p_2-q)^2=k^2-s-t$.

\item 
The cross sections involving gluons $g$ in the initial state are given by
$$    \frac{d\sigma(qg\to q\branon\branon)}{dt\,d(k^2)} =
\frac{d\sigma(\bar{q}g\to \bar{q}\branon\branon)}{dt\,d(k^2)} =
\sigma_0 \frac{\alpha_{3}}{4\pi} F_{qg},\qquad
\frac{d\sigma(gg\to g\branon\branon)}{dt\,d(k^2)} =
\sigma_0 \frac{\alpha_3}{4\pi} F_{gg}$$

\end{enumerate}
The $F$ functions are
\begin{eqnsystem}{sys:f}
F_{e\bar{e}}&=& 4 (1-x)^2\left[\frac{x}{2}(3 - 3x + 2x^2) - \frac{x^3}{2}\sin^2
\theta  +
\frac{(1-x)( 1+(1-x)^2 ) }{x\sin^2 \theta}\right]\\
F_{q\bar{q}}&=&
{2 k^4 ( k^4 + s^2 - 2 k^2 t + 2 t ( s + t )  )  ( k^2 ( s + 4 t )-4 t ( s + t ) ) }/{(s^6 t u) }\\
F_{qg} &=&
-{k^4 ( 2 k^4 + s^2 + t^2 - 2 k^2 ( s + t )  )  ( k^4 + 4 s t - k^2 ( s + t )  ) }/{(3 s^6 t u) }\\
F_{gg} &=& {6 k^4 ( k^8 - 2 k^6 ( s + t )  + 3 k^4 ( s^2 + t^2 )  + ( s^2 + s t + t^2 )^2 - 2 k^2 ( s^3 + t^3 )  ) }/{(s^6 t u) }
\end{eqnsystem}

\frenchspacing
\small\footnotesize

\begin{multicols}{2}

\end{multicols}


\begin{thebibliography}{nn}


\bibitem{DDG} 
\art[hep-ph/9803315]{N. Arkani-Hamed, S. Dimopoulos and G. Dvali}{\PL}{B429}{263}{1998};
\art[hep-ph/9804398]{I. Antoniadis, N. Arkani-Hamed, S. Dimopoulos and G. Dvali}{\PL}{B436}{263}{1998}.


\bibitem{nuR5d}
S. Dimopoulos, talk given at the SUSY 1998 conference (July 1998);
\art[hep-ph/9811428]{K.R. Dienes, E. Dudas, T. Gherghetta}{\NP}{B557}{25}{1999};
\hepart[hep-ph/9811448]{N. Arkani-Hamed et alii}.



\bibitem{SN1987+graviton}
\art{S. Cullen and M. Perelstein}{\PRL}{83}{268}{1999};
\hepart[nucl-th/0007016]{C. Hanhart et al.}.

\bibitem{SN1987+nu}
\hepart[hep-ph/0002199]{R. Barbieri, P. Creminelli and A. Strumia}.


\bibitem{GRW}
\art[hep-ph/9811291]{G. Giudice, R. Rattazzi and J. D. Wells}{\NP}{B544}{3}{1999};
\art[hep-ph/9811337]{E.A. Mirabelli, M. Perelstein and M.E. Peskin}{\PRL}{82}{2236}{1999};
\art[hep-ph/9811350]{T. Han, J.D. Lykken and R. Zhang}{\PR}{D59}{105006}{1999};
\art[hep-th/9809147]{Z. Kakushadze and S. H. Tye}{\NP}{B548}{180}{1999};
\art[hep-ph/9904262]{G. Shiu, R. Shrock and S. H. Tye}{\PL}{B458}{274}{1999}.

\bibitem{Lbrane}
See e.g. J. Polchinski, {\em String theory}, Cambridge Univ. press;
\art{R. Sundrum}{\PR}{D59}{085009}{1999}.

\bibitem{Kugo}
\hepart[hep-ph/9912496]{T. Kugo and K. Yoshioka}.
The parameter $f$ employed by Kugo and Yoshioka equals $(2\pi)^{1/2} \delta^{1/8}\times$ ($f$ in this paper).


\bibitem{Lbrane3}
\hepart[hep-ph/0007100]{A. Dobado and A.L. Maroto}.

\bibitem{fat}
\art[hep-ph/0001335]{A. De Rujula, A. Donini, M.B. Gavela, S. Rigolin}{\PL}{B482}{195}{2000}.


\bibitem{gravitini}
\hepart[hep-ph/9801329]{A. Brignole, F. Feruglio, M. Mangano, F. Zwirner}.




\bibitem{L3} \art{The L3 collaboration}{\PL}{B464}{135}{1999}.


\bibitem{D0} \hepart[hep-ph/0008065]{The D0 collaboration}.

\bibitem{recoil}
\art[hep-ph/9906549]{M. Bando, T. Kugo, T. Noguchi and K. Yoshioka}{\PRL}{83}{3601}{1999}.


\bibitem{NRO} \art[hep-ph/9905281]{R. Barbieri, A. Strumia}{\PL}{B462}{144}{1999}.



\bibitem{BraneModels}
\art[hep-ph/0004214]{I. Antoniadis, E. Kiritsis, T.N. Tomaras}{\PL}{B486}{186}{2000};
\hepart[hep-th/0005067]{G. Aldazabal, L.E. Ib\'a\~nez, F. Quevedo, A.M. Uranga}.



\bibitem{OPAL}
\hepart[hep-ex/0005002]{The OPAL collaboration}.
Comparable bounds have been produced by the
DELPHI collaboration, preprint CERN--EP/2000-021,
and by the L3 collaboration, ref.~\cite{L3}.



\bibitem{CompHEP}
CompHEP --- a package for evaluation of Feynman diagrams and integration over multi-particle phase space.
\hepart[hep-ph/9908288]{A.Pukhov, E.Boos, M.Dubinin,
V.Edneral, V.Ilyin, D.Kovalenko, A.Kryukov, V.Savrin, S.Shichanin, A.Semenov}.


\bibitem{500GeV} See e.g.
\hepart[hep-ex/0007022]{The American Linear Collider Working Group}.

\bibitem{partons} We have used the partons distribution functions of
\art{A.D. Martin et al.}{Eur. Phys. J. C}{4}{463}{1998}
and those of
\hepart[hep-ph/9903282]{H.L. Lai et al.}
as published on the http address
{\tt durpdg.dur.ac.uk}.

\end{thebibliography}
\end{document}